# ACP-ESM: A novel framework for classification of anticancer peptides using protein-oriented transformer approach

Zeynep Hilal Kilimci, Mustafa Yalcin*

*Department of Information Systems Engineering, Kocaeli University, 41001, Kocaeli, Turkey*

ABSTRACT

Anticancer peptides (ACPs) are a class of molecules that have gained significant attention in the field of cancer research and therapy. ACPs are short chains of amino acids, the building blocks of proteins, and they possess the ability to selectively target and kill cancer cells. One of the key advantages of ACPs is their ability to selectively target cancer cells while sparing healthy cells to a greater extent. This selectivity is often attributed to differences in the surface properties of cancer cells compared to normal cells. That is why ACPs are being investigated as potential candidates for cancer therapy. ACPs may be used alone or in combination with other treatment modalities like chemotherapy and radiation therapy. While ACPs hold promise as a novel approach to cancer treatment, there are challenges to overcome, including optimizing their stability, improving selectivity, and enhancing their delivery to cancer cells, continuous increasing in number of peptide sequences, developing a reliable and precise prediction model. In this work, we propose an efficient transformer-based framework to identify anticancer peptides for by performing accurate a reliable and precise prediction model. For this purpose, four different transformer models, namely ESM, ProtBert, BioBERT, and SciBERT are employed to detect anticancer peptides from amino acid sequences. To demonstrate the contribution of the proposed framework, extensive experiments are carried on widely-used datasets in the literature, two versions of AntiCp2, cACP-DeepGram, ACP-740. Experiment results show the usage of proposed model enhances classification accuracy when compared to the state-of-the-art studies. The proposed framework, ESM, exhibits 96.45% of accuracy for AntiCp2 dataset, 97.66% of accuracy for cACP-DeepGram dataset, and 88.51 % of accuracy for ACP-740 dataset, thence determining new state-of-the-art.

## 1. Introduction

Cancer is a formidable global health challenge, characterized by the uncontrolled proliferation of abnormal cells within an organism. Its ubiquity on a worldwide scale renders it a significant cause of morbidity and mortality, with substantial implications for public health, healthcare systems, and socioeconomic well-being. Epidemiological data underscores the pervasive nature of cancer, with a remarkable incidence across diverse populations and geographic regions. Surveillance efforts, such as those led by the World Health Organization (WHO) and national cancer registries, have consistently reported escalating figures. As of the most recent available data, an estimated 19.3 million new cancer cases were diagnosed globally in 2020, marking a continuous upward trend. Cancer manifests as a heterogeneous group of diseases, characterized by the affected organ or tissue, cellular origin, and genetic alterations. This heterogeneity results in the existence of more than 100 distinct cancer types, each endowed with specific clinical and molecular features. Prominent among these are breast, lung, colorectal, prostate, and skin cancers. These entities exhibit varying incidence rates and clinical behaviors, necessitating tailored diagnostic and therapeutic approaches.

Traditional cancer treatments, including chemotherapy and radiation therapy, frequently come with limitations concerning their effectiveness, specificity, and the presence of undesirable side effects. Conventional cancer treatments like chemotherapy and radiation therapy have been widely used and have demonstrated effectiveness in treating cancer. However, their efficacy can vary depending on factors such as the type and stage of cancer and the individual patient's response. Another limitation of these treatments is their lack of specificity. Chemotherapy, for example, targets rapidly dividing cells, which includes cancer cells, but it can also affect healthy cells in the body. This lack of specificity can lead to side effects. Both chemotherapy and radiation therapy can indeed result in significant side effects. Chemotherapy can cause nausea, vomiting, fatigue, hair loss, and anemia, among other issues. Radiation therapy can lead to skin irritation and damage to nearby healthy tissues and organs, depending on the treatment site. These side effects can have a substantial impact on a patient's quality of life.

*Corresponding author: Zeynep Hilal Kilimci
✉ zeynep.kilimci@kocaeli.edu.tr (Z.H. Kilimci); mstf.yalcin@outlook.com (M. Yalcin)
ORCID(s):

On the other hand, peptide-based therapies, specifically anticancer peptides (ACPs), represent a unique and promising area of research in cancer treatment. While ACPs are not considered conventional cancer therapies like chemotherapy and radiation, they have gained attention for their potential advantages and are sometimes explored as alternative or complementary options. ACPs can be designed to specifically target cancer cells by interacting with molecules or receptors that are overexpressed or unique to those cells. This high specificity reduces the risk of harming healthy cells, making ACPs a potentially selective and precise treatment option. Compared to some traditional cancer treatments, ACPs often exhibit lower toxicity profiles. Their selectivity for cancer cells can result in fewer side effects and less damage to normal tissues, thereby improving the patient's quality of life during treatment. ACPs can have diverse mechanisms of action, such as disrupting cancer cell membranes, interfering with signaling pathways, inducing apoptosis (cell death), or modulating the immune response. This versatility allows for the development of ACPs tailored to different types of cancers and molecular characteristics. ACPs may offer an advantage in addressing resistance mechanisms that cancer cells can develop against traditional chemotherapeutic agents. By targeting alternative pathways, ACPs may overcome resistance and remain effective in treating resistant tumors. Peptide-based therapies can be customized based on the specific molecular profile of a patient's cancer. This aligns with the principles of precision medicine, where treatment is tailored to the individual's unique genetic and molecular characteristics. ACPs can be used in combination with other cancer treatments, such as chemotherapy or radiation therapy, to enhance treatment outcomes. This synergistic approach may improve therapeutic efficacy and reduce the risk of relapse. Peptide-based vaccines can stimulate the patient's immune system to recognize and target cancer cells. This immunotherapeutic approach holds promise in harnessing the body's natural defenses to fight cancer.

In this work, we propose a novel framework for classification of ACPs using protein-based transformer approach, called ACP-ESM, especially considering the advantages of ACPs in cancer treatment mentioned above. To achieve this objective, we employ four distinct transformer models: ESM, ProtBert, BioBERT, and SciBERT, for the purpose of identifying anticancer peptides within amino acid sequences. To showcase the effectiveness of our proposed framework, we conduct comprehensive experiments on well-established datasets commonly referenced in the literature, namely AntiCp2, Independent, MRZ, and iACP. The experimental outcomes clearly indicate that the utilization of our proposed model leads to a significant enhancement in classification accuracy compared to the existing state-of-the-art studies. Specifically, our proposed framework, ESM, achieves an accuracy of 96.45% for the AntiCp2 dataset, 97.66% for the cACP-DeepGram dataset, and 88.51% for the ACP-740 dataset, consequently establishing a new state-of-the-art benchmark.

The main contributions of the paper are as follows:

• Introduction of a novel transformer-based framework designed to identify anticancer peptides accurately, showcasing the potential of advanced deep learning techniques in this domain.

• Comprehensive evaluation using four different transformer models, ESM, ProtBert, BioBERT, and SciBERT, demonstrating the versatility of the proposed framework and allowing for a comparative analysis of model performance.

• Establishment of new benchmarks in anticancer peptide prediction by achieving state-of-the-art accuracy rates, particularly with the ESM model, which outperformed existing studies on popular datasets.

• Robust performance on widely-used datasets in the literature, including AntiCp2, cACP-DeepGram, and ACP-740, indicating the generalizability and adaptability of the proposed framework.

• Achieving remarkable accuracy rates while maintaining a low level of complexity with the ESM model which ensures optimal performance without unnecessary computational burden.

• Exhibiting notably low inference times with the ESM model, making it efficient for quick predictions and real-time applications in anticancer peptide identification.

The subsequent sections of this paper are organized as follows: Section 2 provides a comprehensive overview of previous research pertaining to the categorization of anticancer peptides. Section 3 and Section 4 offer a detailed exposition of the proposed framework and transformer models used in the study. Section 5 and Section 6 delineate the findings of our experiments and present the concluding remarks, respectively.

## 2. Literature Review

In this section, literature studies on anticancer peptides (ACPs) classification are briefly introduced.

In (Chen et al., 2021), a deep learning approach rooted in convolutional neural networks is introduced for the prediction of biological activity, specifically focusing on EC50, LC50, IC50, and LD50 values pertaining to six distinct tumor cell types, which encompass breast, colon, cervix, lung, skin, and prostate cells. The research underscores the superior performance of models crafted through multitask learning as compared to conventional single-task models. The comprehensive evaluation of these models is conducted through a series of repeated 5-fold cross-validation exercises, employing the CancerPPD dataset. The results reveal that the most adept models, defined within the confines of an applicability domain, yielded an average mean squared error of 0.1758, a Pearson's correlation coefficient of 0.8086, and a Kendall's correlation coefficient of 0.6156. The culmination of these efforts has yielded a novel method termed "xDeep-AcPEP," which is anticipated to prove invaluable in the realm of rational peptide design for therapeutic purposes. Specifically, this approach holds the potential to identify and characterize effective ACPs, thus advancing the field of peptide-based therapeutics.

In (Nasiri et al., 2021), an assortment of machine learning algorithms, specifically Random Forest (RF), Support Vector Machine (SVM), and eXtreme Gradient Boosting (XGBoost), has been employed to undertake the prediction of active Cp-ACPs, utilizing a meticulously validated experimental dataset. A novel model, denoted as CpACpP, has been meticulously devised, grounded in the amalgamation of two distinct subpredictors, one focused on Cell-Penetrating Peptides (CPP) and the other on ACPs, both functioning independently. To augment the predictive capacity of the model, a diverse array of compositional and physiochemical-based features has been strategically amalgamated or selectively curated, leveraging the multilayered Recursive Feature Elimination (RFE) methodology, spanning both datasets. The outcomes of the analysis reveal that the ACP subclassifiers consistently achieve a commendable mean performance accuracy (ACC) standing at 0.98, accompanied by an area under the curve (AUC) approximate to 0.98. In contrast, the CPP predictors exhibit closely aligned performance metrics, with values approximating 0.94 and 0.95, respectively, when contemplating the hybrid-based features and the independent data sets.

A novel deep-learning-based predictor, referred to as ACPred-LAF, is introduced in the study (He et al., 2021). Within this framework, a multisensory and multiscaled embedding algorithm is proposed to autonomously acquire and extract contextual sequential attributes associated with ACPs. Comprehensive feature comparisons establish the superiority of our learnable and self-adaptive embedding features over manually crafted ones in capturing distinctive information, thereby substantially enhancing the predictive accuracy for ACPs. Furthermore, benchmarking comparisons conduct on both established benchmark datasets and a newly curated dataset affirm the superior performance of ACPred-LAF when compared to existing state-of-the-art methods. To fortify the credibility and resilience of the model, a thorough investigation is conducted through a data interference experiment, validating its robustness under varying conditions.

In (Wan et al., 2021), two distinct models are constructed with the aim of discriminating Anticancer Peptides (ACPs) from antimicrobial peptides (AMPs) and predicting ACPs from the entirety of peptide sequences. The selection of features encompasses amino acid composition, N5C5, k-space, and position-specific scoring matrix (PSSM). Machine learning methods, including Support Vector Machine (SVM) and Sequential Minimal Optimization (SMO), are employed to develop Model 2, which specializes in distinguishing ACPs from complete peptides. Additionally, Model 1 is designed to differentiate ACPs from AMPs. Notably, when compared to preceding models, the models devised in this research exhibit superior performance, achieving accuracy rates of 85.5% for Model 1 and 95.2% for Model 2.

n (Ahmed et al., 2021), a novel multi-headed deep convolutional neural network model, denominated as ACP-MHCNN, is introduced. The primary objective of this model is to effectively extract and integrate discriminative features from diverse information sources in an interactive manner. ACP-MHCNN achieves this by extracting sequence, physicochemical, and evolutionary-based features for the identification of anticancer peptides (ACPs). This process involves the utilization of various numerical peptide representations, all while minimizing parameter overhead. Through a series of extensive experiments, encompassing cross-validation and an independent dataset, it becomes apparent that ACP-MHCNN exhibits a significant performance advantage over other models employed for the identification of anticancer peptides. In comparison to state-of-the-art models, ACP-MHCNN showcases substantial improvements, boasting higher accuracy, sensitivity, specificity, precision, and MCC rates by margins of 6.3%, 8.6%, 3.7%, 4.0%, and 0.20, respectively.

In (Park et al., 2022), the researchers establish the initial comprehensive and non-repetitive training and independent datasets for the investigation of ACPs. Within the confines of the training dataset, a broad spectrum of feature encoding methods is explored, leading to the development of distinct models using seven different conventional classifiers. Following this, a subset of encoding-based models exhibiting superior performance for each classifier is chosen, and their predicted scores are amalgamated and subsequently trained through a convolutional neural network (CNN), thereby yielding a predictor referred to as MLACP 2.0. The assessment of MLACP 2.0, using a notably diverse independent dataset, demonstrates exceptional performance, surpassing recent tools designed for ACP prediction. Furthermore, MLACP 2.0 displays heightened effectiveness in both cross-validation and independent evaluation in comparison to CNN-based embedding models and traditional single models. .

In (Sun et al., 2022), a deep learning-based model, denoted as ACPNet, is introduced for the discrimination of anticancer peptides from non-anticancer peptides (non-ACPs). ACPNet incorporates three distinct categories of peptide sequence information, encompassing peptide physicochemical properties and auto-encoding features that are integrated into the training process. ACPNet represents a hybrid deep learning network that amalgamates fully connected networks with recurrent neural networks. Through comparisons with existing methods utilizing the ACPs82 dataset, it is evident that ACPNet not only attains a 1.2% improvement in accuracy but also showcases enhancements of 2.0% in F1-score, 7.2% in recall, and demonstrates balanced performance in the Matthews correlation coefficient. Furthermore, an independent validation experiment was conducted, involving a dataset comprising 20 established anticancer peptides, where ACPNet impressively identified only one anticancer peptide as a non-ACP. These comparative analyses and the independent validation experiment collectively affirm the precise discriminative capabilities of ACPNet in distinguishing anticancer peptides from non-ACPs.

A dependable framework has been devised to facilitate the precise identification of ACPs in (Alsanea et al., 2022). This approach incorporates four distinct hypothetical feature encoding mechanisms, which encompass amino acid, dipeptide, tripeptide, and an enhanced iteration of pseudo-amino acid composition, each serving as indicators of the motif characteristic to the target class. To streamline feature selection, Principal Component Analysis (PCA) is harnessed, enabling the identification of optimal, profound, and highly diversified features. Given the inherent variability in learning techniques, a battery of experiments is conducted involving various algorithms to discern the most effective operational method. Following a thorough examination of empirical outcomes, it becomes evident that the Support Vector Machine, when applied in conjunction with the hybrid feature space, demonstrates superior performance. As a result of these developments, the proposed framework attains an impressive accuracy rate of 97.09% and 98.25% on benchmark and independent datasets, respectively. A comparative analysis underscores the superior performance of our proposed model when juxtaposed with existing methods, signifying its potential significance in the realms of drug development and oncology.

In (Ghulam et al., 2022), a novel deep learning-based approach, denoted as ACP-2DCNN, is introduced with the primary objective of enhancing the predictive accuracy for ACPs. During the process of model training and prediction, pivotal features are extracted utilizing the Dipeptide Deviation from Expected Mean (DDE) methodology. Subsequently, Two-dimensional Convolutional Neural Network (2D CNN) techniques are employed. The empirical findings emanating from this study underscore the efficacy of the proposed method, which has surpassed existing methodologies in the literature by demonstrating superior performance. Specifically, ACP-2DCNN exhibits enhanced accuracy in ACP prediction when compared to its predecessors.

In (Akbar et al., 2022), a word embedding strategy based on FastText has been deployed to represent individual peptide samples, leveraging a skip-gram model. Following the derivation of peptide embedding descriptors, a deep neural network (DNN) model is subsequently employed to carry out precise discrimination of ACPs. The optimization of DNN model parameters has yielded an impressive accuracy rate of 96.94% for the training dataset, 93.41% for alternate samples, and 94.02% for independent samples, respectively. Notably, a comparative assessment reveals the superior performance of our proposed cACP-DeepGram model, demonstrating a notable increase of approximately 10% in predictive accuracy when contrasted with existing predictive models.

Multiple predictive models have been developed within this research Alimirzaei and Kieslich (2023), employing a spectrum of machine learning techniques, including Support Vector Machines (SVMs), Gradient Boosting Classifiers (GB), and Random Forest Classifiers (RF). The focal objective is the anticipation of membranolytic anticancer activity based on peptide sequences. Notably, this study leverages the established correlation between oscillations in physicochemical properties within protein sequences and predictive insights into protein structure and function. To achieve this objective, Fourier transforms have been systematically applied to property factor vectors, facilitating the quantification of the amplitude of physicochemical oscillations. These amplitude measurements serve as pivotal features within the predictive models. The dataset encompasses peptides specifically targeting breast and lung cancer cells, culled from the CancerPPD database. These peptides are subsequently translated into physiochemical vectors through the utilization of 10 distinct property factors, corresponding to the 20 natural amino acids. The predictive models undergo a rigorous process of training and refinement through cross-validation, involving the iterative division of the dataset into multiple training and testing subsets. Feature selection techniques are judiciously applied to further enhance the efficiency of SVM models. The evaluation of model performance hinges upon the quantification of cross-validation classification accuracy. Furthermore, in the quest to refine prediction accuracy, this study extends its purview to consider alternative sets of physiochemical features and amino acid properties, drawn from relevant literature sources, as complementary attributes incorporated into the predictive models.

In (Deng et al., 2023), a computational tool termed ACP-MLC has been introduced to tackle the binary classification and multi-label classification of ACPs based on given peptide sequences. ACP-MLC operates as a two-tiered prediction system: in the first level, it employs the random forest algorithm to predict whether a given sequence qualifies as an ACP. In the second level, it employs the binary relevance algorithm to predict which tissue types the sequence may target. Following development and rigorous evaluation using high-quality datasets, ACP-MLC achieves notable results. For the first-level prediction, it attains an area under the receiver operating characteristic curve (AUC) of 0.888 on the independent test set. For the second-level prediction, it records a hamming loss of 0.157, subset accuracy of 0.577, F1-score macro of 0.802, and F1-score micro of 0.826 on the independent test set. A comprehensive comparative analysis reveals that ACP-MLC outperforms existing binary classifiers and other multi-label learning classifiers in the context of ACP prediction.

In (Fazal et al., 2023), an alternative classification approach rooted in the established statistical theory of sparse-representation classification (SRC) is introduced, diverging from the conventional black-box methodologies. The approach revolves around the construction of over-complete dictionary matrices through the incorporation of K-spaced amino acid pairs (CKSAAP). Differing from traditional SRC frameworks, this strategy employs an efficient matching pursuit solver in lieu of the computationally intensive basis pursuit solver. Moreover, it leverages kernel principal component analysis (KPCA) to address non-linearity and reduce dimensionality within the feature space. Additionally, the synthetic minority oversampling technique (SMOTE) is deployed to balance the dictionary. The proposed methodology undergoes rigorous assessment using two benchmark datasets, and its performance is gauged against established statistical parameters. The outcomes reveal that this method surpasses existing approaches, exhibiting notably heightened sensitivity and superior balanced accuracy.

A novel machine learning framework, denoted as GRDF, has been introduced for the purpose of ACP identification in (Yao et al., 2023). GRDF incorporates deep graphical representation techniques and a deep forest architecture, amalgamating these components to harness the discriminative power required for ACP discrimination. More specifically, GRDF operates by extracting graphical features founded on the physicochemical attributes inherent to peptides. These features are further enriched through the integration of evolutionary information and binary profiles to construct robust models. Additionally, the framework employs the deep forest algorithm, characterized by its layer-by-layer cascade architecture, reminiscent of deep neural networks. This architectural choice allows for exemplary performance even with limited dataset sizes, eliminating the need for intricate hyperparameter tuning. Experimental outcomes underscore the effectiveness of GRDF, as it attains state-of-the-art results on two comprehensive datasets, referred to as Set 1 and Set 2. It achieves an accuracy of 77.12% and an F1-score of 77.54% on Set 1, while obtaining an accuracy of 94.10% and an F1-score of 94.15% on Set 2, thus surpassing existing methods in ACP prediction.

In (Yang et al., 2023), Contrastive ACP Predictor (CACPP) is introduced as a deep learning framework that leverages convolutional neural networks (CNN) and employs the principles of contrastive learning to achieve precise predictions of anticancer peptides. Specifically, the TextCNN model is harnessed to derive high-latent features exclusively from the peptide sequences. Furthermore, the contrastive learning module is incorporated to facilitate the acquisition of more discernible feature representations, thereby enhancing the prediction accuracy. Empirical findings gleaned from comprehensive evaluations on benchmark datasets unequivocally demonstrate that CACPP surpasses all existing state-of-the-art methodologies in the realm of anticancer peptide prediction.

An enhanced predictor for anticancer peptides, denoted as ACP-GBDT, has been introduced in (Li et al., 2023). This predictor hinges on the utilization of gradient boosting decision tree (GBDT) techniques and leverages sequence information to bolster its predictive capabilities. For the purpose of encoding the peptide sequences encompassed within the anticancer peptide dataset, ACP-GBDT employs a merged-feature approach, which combines information sourced from AAIndex and SVMProt-188D. The training of the prediction model within ACP-GBDT is executed via the GBDT framework. Through rigorous evaluation encompassing independent testing and ten-fold cross-validation, it has been established that ACP-GBDT is adept at effectively discriminating anticancer peptides from their non-anticancer counterparts. Furthermore, comparative analyses conducted on benchmark datasets have demonstrated that ACP-GBDT excels in both simplicity and effectiveness when juxtaposed with existing methods for anticancer peptide prediction.

## 3. Models

In this section, a comprehensive overview is presented of transformer models that have been specifically devised and further developed for the purpose of ACP classification.

### 3.1. Evolutionary scale modeling (ESM)

ESM (Lin et al., 2023) serves as a specialized framework within the field of computational biology and bioinformatics, designed to enhance understanding of proteins, their functions, and structures. It provides improved protein analysis, biological discovery, and functional annotation. ESM aims to provide more accurate and comprehensive insights into proteins by leveraging large-scale sequence data and evolutionary information. It enhances ability to analyze and predict various aspects of proteins, including their structures, functions, and interactions. ESM facilitates the discovery of novel patterns, relationships, and features within protein sequences. By considering the evolutionary history of proteins, it contributes to understanding of how proteins have evolved to perform specific biological roles. Furthermore, ESM assists in the annotation of protein functions. It can predict the functions of uncharacterized proteins based on similarities with known proteins, helping researchers identify potential roles in biological processes.

The structure of ESM involves several key components and methodologies.

**i) Pretraining:** ESM models are pretrained on large datasets containing protein sequences. This pretraining phase often employs deep learning architectures, similar to those used in natural language processing. During pretraining, the model learns to capture intricate patterns and representations from the protein sequences.

**ii) Evolutionary Information:** A fundamental aspect of ESM is its incorporation of evolutionary information. This is achieved by analyzing multiple sequence alignments (MSAs) of related proteins. MSAs provide insights into the variations and changes that have occurred in protein sequences over time. ESM models use this evolutionary history to capture dependencies and evolutionary signals.

**iii) Fine-Tuning:** After pretraining, ESM models are fine-tuned for specific protein-related tasks. This involves adapting the pretrained model to perform tasks such as protein function prediction, protein-protein interaction prediction, or structural analysis. Fine-tuning allows the model to specialize in these tasks and make accurate predictions.

**iv) Multiple Sequence Alignments (MSAs):** MSAs play a crucial role in ESM. They are collections of related protein sequences, often obtained from various sources. MSAs provide a wealth of information about how proteins have evolved and help ESM models learn from the diverse characteristics of protein families.

**v) Deep Learning Architectures::** ESM models are typically built using deep learning architectures, such as recurrent neural networks (RNNs) or transformer-based models. These architectures are capable of capturing complex relationships and patterns within protein sequences.

In summary, ESM is a sophisticated framework tailored for protein analysis. It combines deep learning techniques with evolutionary insights to advance our understanding of proteins' roles in biological systems. By focusing on large-scale sequence data and evolutionary history, ESM contributes to various bioinformatics applications and enables researchers to uncover the mysteries of proteins at a molecular level.

### 3.2. A pretrained model on protein sequences (ProtBert)

ProtBERT (Elnaggar et al., 2021) is a specialized variant of the BERT (Bidirectional Encoder Representations from Transformers) model Kenton and Toutanova (2019) that is specifically designed for understanding and processing protein sequences and biological data. Like BERT, which was originally developed for natural language understanding, ProtBERT is pre-trained on a large corpus of protein sequences and then fine-tuned for specific downstream tasks in the field of computational biology. This pre-trained model, ProtBERT, captures the contextual information and representations of amino acids in protein sequences, allowing it to understand the relationships, structures, and functions of proteins more effectively. It has shown promise in various bioinformatics tasks, including protein classification, protein-protein interaction prediction, and protein function prediction. ProtBERT is an example of how transfer learning techniques, originally developed for natural language processing, can be adapted and fine-tuned for specialized domains like computational biology to improve the understanding and analysis of biological data.

Just like BERT, ProtBERT tokenizes input sequences into smaller units. In the case of natural language, these are usually words or subwords, while for proteins, the tokens represent amino acids. ProtBERT uses a tokenization scheme suitable for protein sequences, typically breaking down amino acid sequences into tokens. Each token is associated with an embedding vector. These vectors are learned during the training process and represent the features of each token. In ProtBERT, these embeddings capture information about the amino acids and their relationships. The core of ProtBERT consists of multiple transformer blocks. Each block contains two main components:

**i) Multi-Head Self-Attention Mechanism:** This mechanism allows the model to weigh the importance of each token in relation to all other tokens in the sequence, capturing dependencies and relationships between amino acids in the protein sequence.

**ii) Positional Encoding:** Since transformers don't inherently consider the order of tokens, positional encodings are added to the embeddings to give the model information about the position of each token in the sequence.

ProtBERT consists of multiple transformer blocks stacked on top of each other. These layers allow the model to learn increasingly abstract and complex features from the input data. In pre-training, ProtBERT is pretrained on a large corpus of protein sequences, similar to how BERT is pretrained on text from the internet. During pre-training, the model learns to predict masked amino acids in protein sequences and, in the process, captures useful representations of the sequences. After pretraining, ProtBERT can be fine-tuned for specific downstream tasks in bioinformatics, such as protein classification, protein-protein interaction prediction, or protein function prediction. Fine-tuning involves training the model on a smaller task-specific dataset, which allows it to specialize in these tasks. ProtBERT typically uses a specialized vocabulary that includes amino acids and amino acid subsequence representations, as opposed to the word-based vocabulary used in BERT for natural language.

Overall, ProtBERT adapts the transformer architecture to the domain of protein sequences, making it a powerful tool for various computational biology tasks by effectively capturing the contextual information, relationships, and features of amino acids in proteins.

### 3.3. A pre-trained biomedical language representation model (BioBERT)

In the capacity of a comprehensive language representation model, BERT Kenton and Toutanova (2019) underwent pre-training on English Wikipedia and BooksCorpus. Nevertheless, biomedical texts encompass a notable array of domain-specific proper nouns (e.g., BRCA1, c.248T>C) and terminologies (e.g., transcriptional, antimicrobial), which predominantly cater to the comprehension of biomedical researchers. Consequently, NLP models engineered for general-purpose language comprehension often yield subpar performance when applied to biomedical text mining undertakings. In response to this challenge, BioBERT (Lee et al., 2020) is introduced as a pre-trained language representation model tailored to the biomedical domain.

Initially, BioBERT is endowed with weights inherited from BERT, which underwent pre-training on corpora from the general domain (English Wikipedia and BooksCorpus). Subsequently, BioBERT undergoes pre-training on corpora specific to the biomedical domain (PubMed abstracts and PMC full-text articles). BioBERT employs WordPiece tokenization (Wu et al., 2016) for tokenization, which effectively mitigates the issue of out-of-vocabulary terms. Through WordPiece tokenization, novel words can be effectively represented by frequently occurring subword components (e.g., Immunoglobulin ¯ I ##mm ##uno ##g ##lo ##bul ##in). It is observed that the utilization of a cased vocabulary (as opposed to lowercase) yields slightly improved performance in downstream tasks. While an alternative approach could have involved the construction of a new WordPiece vocabulary based on biomedical corpora, the original vocabulary from BERTBASE is employed for two primary reasons:

• To ensure BioBERT's compatibility with BERT, facilitating the reuse of BERT pre-trained on general domain corpora and promoting the seamless interchangeability of models based on BERT and BioBERT;

• Any novel terms could still be adequately represented and fine-tuned for the biomedical domain using BERT's original WordPiece vocabulary.

To underscore the effectiveness of BioBERT in biomedical text mining, BioBERT is subjected to fine-tuning and evaluation across three prominent biomedical text mining tasks (NER, RE, and QA). Various pre-training strategies, encompassing diverse combinations and scales of general domain corpora and biomedical corpora, are tested, and the influence of each corpus on pre-training is meticulously analyzed.

### 3.4. A pretrained language model for scientific text (SciBERT)

SciBERT (Beltagy et al., 2019) is a language model pretrained on BERT Kenton and Toutanova (2019) to address the scarcity of high-quality, large-scale annotated scientific data. It achieves this by utilizing unsupervised pre-training on an extensive, multi-domain collection of scientific publications, enhancing its performance in subsequent scientific Natural Language Processing (NLP) tasks. A comprehensive assessment of SciBERT encompasses various tasks, including sequence tagging, sentence classification, and dependency parsing, employing datasets from diverse scientific fields. Statistically significant enhancements over BERT are demonstrated, and new state-of-the-art results are achieved in several of these tasks. SciBERT is trained on a random sample of 1.14 million papers from Semantic Scholar (Ammar et al., 2018). This dataset comprises 18% computer science papers and 82% biomedical papers from a wide range of sources. The entire paper content, rather than just abstracts, is used in the training process. On average, each paper consists of 154 sentences (equivalent to 2,769 tokens), resulting in a substantial corpus of 3.17 billion tokens, which closely matches the scale of BERT's training data (3.3 billion tokens). Sentence splitting is carried out using ScispaCy (Neumann et al., 2019), a tool optimized for scientific text segmentation.

In summary, SciBERT, a novel resource, is introduced, demonstrating its capacity to enhance performance across various NLP tasks within the scientific domain. SCIBERT is a pretrained language model that builds upon BERT but is specifically trained on an extensive corpus of scientific text. Extensive experiments are conducted to explore the performance implications of fine-tuning versus employing task-specific architectures on top of frozen embeddings, along with the impact of domain-specific vocabulary. SCIBERT's effectiveness is evaluated across a spectrum of scientific domain tasks, achieving new state-of-the-art results in numerous instances.

## 4. Proposed Framework
### 4.1. Dataset

In the AntiCp2 Lv et al. (2021) study, datasets termed as the main and alternate datasets have been established. ACPs have been acquired from prior study datasets encompassing ACP-DL, ACPP, ACPred-FL, AntiCP, and iACP. The alternate dataset incorporates both anticancer and random peptides; the assumption being that random peptides are non-ACPs. For the purpose of establishing a balanced dataset, 970 random peptides have been selected and designated as non-ACPs. In simpler terms, the alternate dataset comprises 970 experimentally validated ACPs and an equal number of non-ACPs (i.e., random peptides). In the present investigation, two distinct datasets are formulated, denoted as the main and alternate datasets. The alternate dataset comprises both anticancer peptides and randomly selected peptides, with the presumption that random peptides lack ACP attributes. Ensuring dataset equilibrium, 970 random peptides are selected and classified as non-ACPs. In essence, the alternate dataset encompasses 970 experimentally validated ACPs and a commensurate number of non-ACPs (or random peptides).

In the realm of deep learning, the acquisition or creation of a valid benchmark dataset assumes significance for the training of computational models. Accordingly, two distinct training datasets are utilized in the cACP-DeepGram Akbar et al. (2022) study. Initially, the LEE samples are employed for model training. The LEE dataset, comprising 844 samples, incorporates 422 ACPs and an equal number of non-ACPs. The LEE samples, utilized for model training, exhibit amino acid residue lengths in each sample that are less than 25. Furthermore, to evaluate the model's generalization ability and ascertain the presence of overfitting issues, an independent dataset labeled "Independent Dataset" is incorporated. Independent encompasses 300 samples, with an equal distribution of 150 positive and 150 negative samples. The "cACP-DeepGram" dataset includes the "Lee dataset" with 422 positive and 422 negative samples, along with an "Independent dataset" comprising 150 positive and 150 negative samples.

The ACP-740 Ahmed et al. (2021) dataset, is introduced through the compilation of 388 experimentally validated ACPs for initial construction. Of these, 138 are sourced from Chen et al. (2016), and 250 are obtained from Wei et al. (2018a). Correspondingly, 456 antimicrobial peptides (AMP) lacking anticancer activity are initially gathered, with 206 originating from Chen et al. (2016) and 250 from Wei et al. (2018a). This process culminates in the establishment of a dataset comprising 740 samples, encompassing 376 positives and 364 negatives.

Table 1: Statistics of the Datasets

| Dataset | Positive | Negative | Total |
|---|---|---|---|
| Anticp2 Alternate | 970 | 970 | 1940 |
| cACP-DeepGram | 572 | 572 | 1144 |
| ACP-740 | 376 | 364 | 740 |

Table 1 provides a detailed statistical overview of the datasets, incorporating dataset names, labels, and their corresponding counts. The table specifically includes counts for each dataset and serves to offer a foundational understanding of the quantity of data available in each set. Table 2 presents an elaborate breakdown of the contents within each dataset, encompassing dataset names, contents, and labels. Table 2 enumerates the variables or features present, their respective data types, and any categorical distinctions.

Table 2: Content of the Datasets

| Dataset | Content | Label |
| --- | --- | --- |
| Anticp2 Alternate | FFGWLIKGAIHAGKAIHGLIHRRRH | 1 |
| Anticp2 Alternate | RKAVLLEEQGIEWKPEDTARPSGPREGGRRDGGRDG | 0 |
| cACP-DeepGram | ACDCRGDCFCGGGGIVRRADRAAVP | 1 |
| cACP-DeepGram | ATAPTTRNLLTTPKF | 0 |
| ACP-740 | GLLDIVKKVVGAFGSL | 1 |
| ACP-740 | MTISLIWGIAMVVCCCIWVIFDRRRRKAGEPPL | 0 |

## 4.2. Proposed Methodology

Aspiring to create an efficient classification framework for anticancer peptides, this research initiative amalgamates advanced transformer model methodologies and systematically explores the distinctive features of anticancer peptides, which are derived from extensively employed datasets. The general flowchart of the proposed method framework is presented in Figure 1. The detailed architecture of the proposed transformer model ESM is demonstrated in Figure 2. By employing transformer model algorithms, namely ESM, ProtBert, BioBERT, and SciBERT, vector representations are generated for each instance within the datasets.

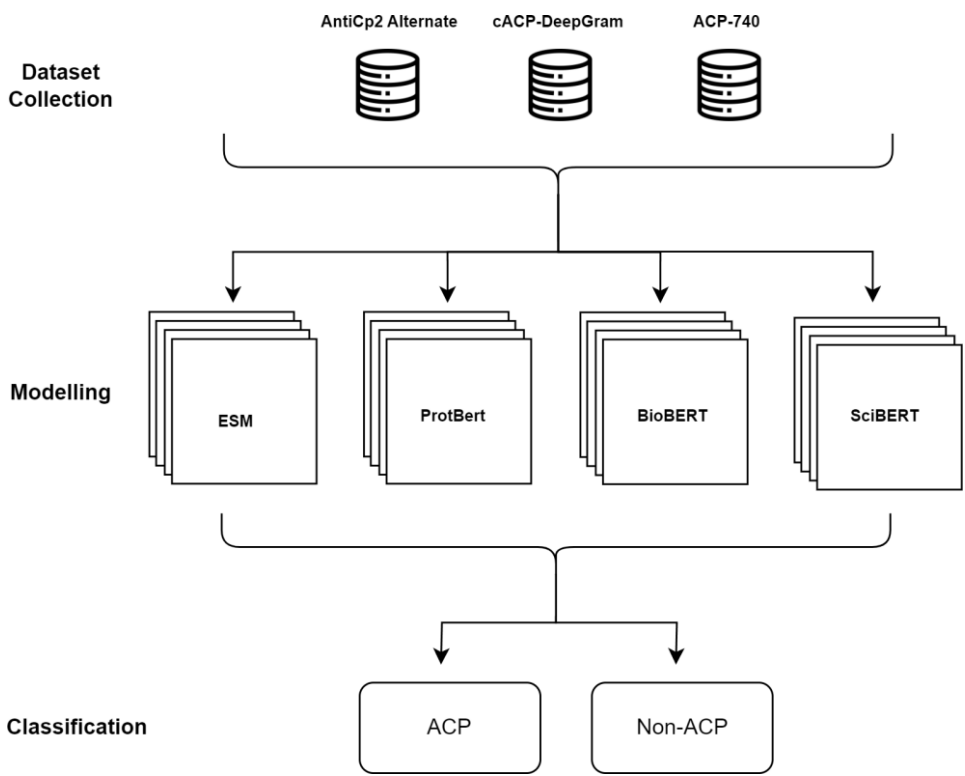

**Figure 1:** A general flow chart of the proposed model

Yearning to achieve an optimized model, our research has initiated a systematic exploration of diverse parameter configurations throughout the model training phase. This comprehensive method ensures a nuanced examination of the performance landscape, considering pivotal parameters such as learning rate, batch size, weight decay, warm-up ratio, and gradient accumulation steps. Effectively addressing the challenge of overfitting, we have implemented a spectrum of learning rates, spanning from 1e-4 to 1e-7. Concurrently, we have meticulously scrutinized batch sizes ranging from 10 to 128, assessing their impact on mitigating concerns related to overfitting. Leveraging the amalgamation of these varied parameter combinations has facilitated a comprehensive analysis of the models' performance. Notably, the study culminates in the identification of optimal models for each dataset, thereby contributing significantly to the realm of anticancer peptide classification. Integral to this study is the strategic application of diverse learning rates between 1e-4 and 1e-7 to alleviate overfitting issues. This approach is posited to empower the model to acquire more generalized features, enhancing adaptability across diverse datasets. Moreover, the exploration of different batch sizes serves as a pivotal step in assessing the scalability and practical applicability of the model.

Largely, the optimization of transformer models mandates meticulous adjustments to hyperparameters, crucial for enhancing performance and mitigating prevalent issues, notably overfitting. Within this purview, the weight decay parameter emerges as a pivotal regulator, exercising control over the magnitudes of model weights. In this context, a systematic modulation of the weight decay parameter within the prescribed interval of 0.01 to 0.0001 efficaciously governs larger weights, thereby attenuating the propensity for overfitting. This nuanced calibration not only alleviates the vulnerability to overfitting but also fortifies the model's generalization capabilities. Leaning into the integration of the warm-up ratio parameter, spanning the judicious range of 0.1 to 0.7, assumes a transformative role in orchestrating learning rate dynamics during the nascent phases of training. The deliberate augmentation of the learning rate at the initiation of training, followed by a gradual escalation, substantially contributes to the model's stability. This adaptive regimen ensures a measured and resilient response to abrupt perturbations during the training trajectory.

Table 3: Parameter Details of the Transformer Models

| Parameters | ESM | ProtBert | BioBERT | SciBERT |
|---|---|---|---|---|
| Hidden Act | gelu | gelu | gelu | gelu |
| Hidden Size | 640 | 1024 | 768 | 768 |
| Initializer Range | 0.02 | 0.02 | 0.02 | 0.02 |
| Intermediate Size | 2560 | 4096 | 3072 | 3072 |
| Layer Norm Eps | 1e-05 | 1e-12 | 1e-12 | 1e-12 |
| Mask Token Id | 32 | - | - | - |
| Max Position Embeddings | 1026 | 40000 | 512 | 512 |
| Num Attention Heads | 20 | 16 | 12 | 12 |
| Num Hidden Layers | 30 | 30 | 12 | 12 |
| Position Embedding Type | rotary | absolute | absolute | absolute |
| Attention Probs Dropout Prob | 0.0 | 0.0 | 0.1 | 0.1 |
| Vocab Size | 33 | 30 | 31090 | 28996 |

Implementing simultaneously, the pivotal parameter of gradient accumulation steps, varying between 1 and 10, facilitates adept management of mini-batches. The aggregation of gradients over a specified number of mini-batches preceding parameter updates not only emulates batch processes but also curtails the frequency of model parameter adjustments. This strategic modulation confers stability during training while concurrently mitigating computational exigencies. Vigorously, for effective monitoring and governance of the training regimen, the logging step parameter, set within the ambit of 1 to 10, is instrumentalized. The judicious logging of model performance post each update step or mini-batch operation enables meticulous scrutiny of training dynamics. This approach not only bequeaths heightened control but also facilitates error analysis, thereby expediting the identification of issues such as overfitting.

Nurturing the scrupulous adjustment of hyperparameters, including weight decay, warm-up ratio, gradient accumulation steps, and logging steps, assumes a cardinal role in optimizing the training process. The synergistic impact of these refinements culminates in a harmoniously balanced model, effectively mitigating the perils of overfitting and elevating the model's generalization prowess to scholarly heights. Unquestionably, the results gleaned from this research underscore the efficacy of the proposed framework in achieving remarkable performance in the classification of anticancer peptides. The meticulous consideration of diverse parameter interactions has led to a nuanced understanding of how these factors collectively impact model performance. Transformer model involves utilizing knowledge acquired in one task for another task. This type of learning often occurs by taking the weights of a pre-trained model and applying the learned information of this model to a different task. Table 3 provides the configuration settings of the utilized models, encompassing the values that dictate the models' operational behavior and learning.

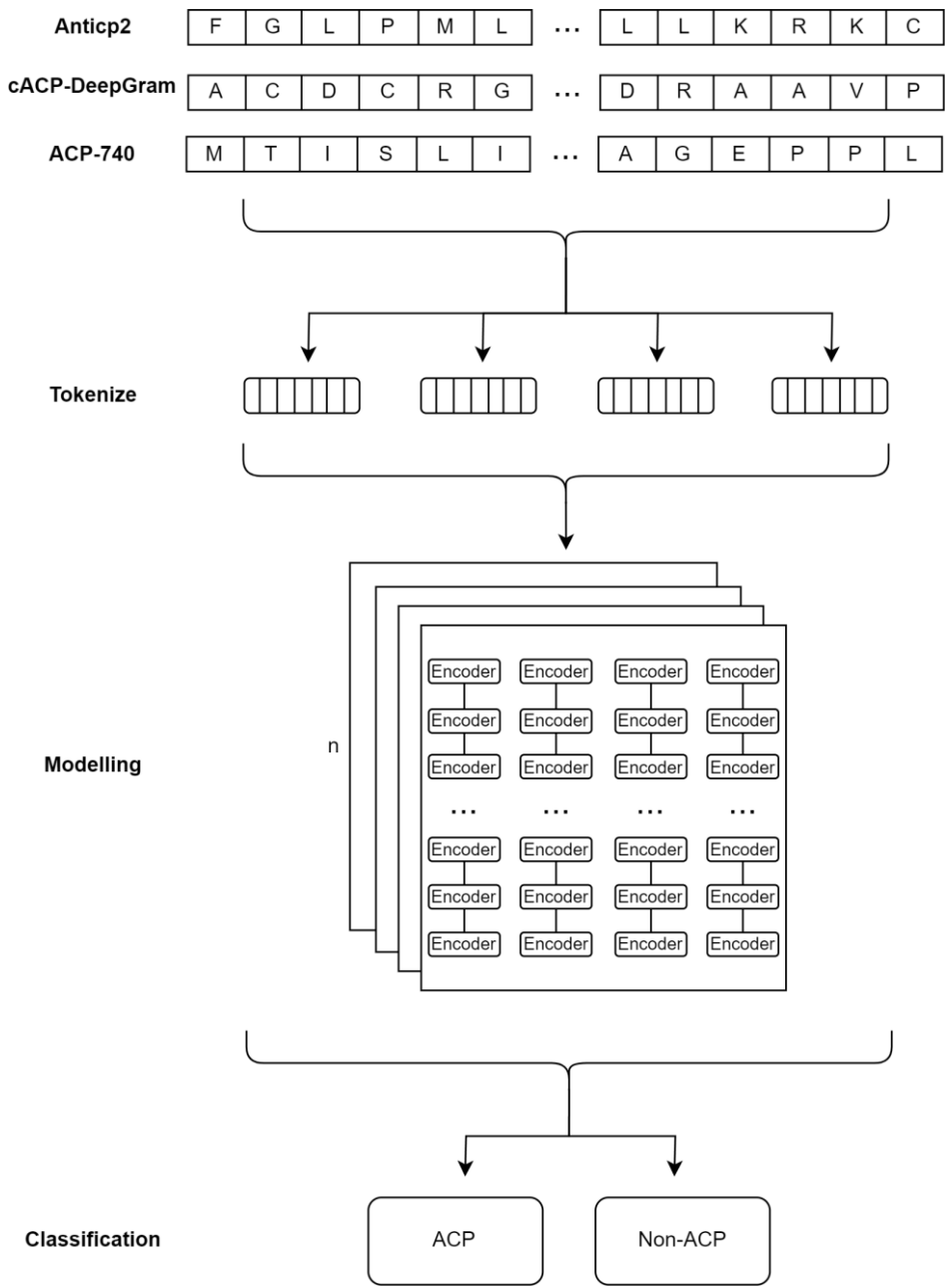

**Figure 2:** The architecture of ESM framework

There are some main components of the architecture of transformers used in the study. These are tokenization, specifying optimization function, and attention map. In tokenization, the input needs to be broke down into small units, known as tokens through tokenizer. Language models rely on these tokens to understand and process words or subwords. The tokenizer converts the text into a numerical form that the model can comprehend. Tokenizing peptide sequences is crucial for understanding the structure of peptides. The model learns the sequence of amino acids in peptides and their interactions, influencing biological effects. This stage includes:

- Representation of Amino Acid Sequences: Peptides are formed by the specific arrangement of amino acids, each usually represented by a symbol. To convert the sequence of amino acids into a numerical form, each amino acid symbol is mapped to a unique numerical value.

- Tokenization of Peptide Sequences: Considering a peptide sequence as a series of amino acid symbols, a tokenizer is employed to transform this sequence into a numerical form understandable by the model. This tokenization process breaks the peptide sequence into smaller units, tokens, and associates each token with a numerical value.

- Utilization of Tokens as Model Input: The numerical values obtained from the tokenization of peptides serve as the input for the model. The model analyzes and learns from these numerical representations to make predictions.

- Representation of Peptides: The model learns the features of peptides and utilizes these features through numerically represented tokens. This process is employed to predict the biological activity or other characteristics of a specific peptide.

Each model features a distinctive tokenizer crafted with consideration for the terminology and structure of its respective field. These customizations enable the models to effectively process and understand domain-specific information, contributing to their performance in specialized tasks. After constructing representation of peptides, the training procedure is initiated using the transformer models summarized below:

- SciBERT: It utilizes the foundational WordPiece tokenization of BERT, but includes a specialized tokenization vocabulary to better understand specific terms and structures in biomedical and scientific texts.

- BioBERT: It builds upon BERT's WordPiece tokenization but incorporates a customized tokenization vocabulary specific to biomedical terminology. This adaptation allows for a more effective handling of specific terms in biomedical texts.

- ProtBert: It employs a unique tokenization process designed to understand amino acids in protein sequences and specialized protein terminology. This customization is intended to better represent protein structures and enhance performance in tasks related to proteins.

- ESM: Its tokenization process is customized for these tasks, ensuring meaningful representation of protein sequences and structural information. This adaptation aims to improve the understanding of protein structures and enhance performance in tasks related to protein folding.

In summary, the process of transformer model and tokenization is vital for applying knowledge from one domain, such as language, to another domain, such as peptides. These techniques enable models to understand and make predictions related to the unique features of peptides.

The other significant stage is specifying the optimizer function which is a mathematical algorithm employed to adjust the parameters of a model during the learning process with the goal of minimizing the error between the model's outputs and the actual values. This function typically utilizes the gradient of a loss function during the parameter update, striving to reduce the amount of error in the model's performance. In essence, the optimizer function employs an iterative approach to find the optimal parameter values for the model, facilitating the convergence of model predictions towards actual values. For each model, the AdamW optimizer function has been employed. AdamW is a modification of the Adam optimizer, focusing on regularizing weights during weight updates, thereby enhancing performance. Moreover, ProtBert, SciBERT, and BioBERT models utilize BertPooler and Tanh as an activation function. On the other hand, the ESM model adopts EsmPooler with Tanh as the activation function. As a significant distinction, the ESM model deviates from the other models by employing the sigmoid function for the contact head. This unique choice in the contact head of the ESM model, using sigmoid activation, sets it apart from ProtBert, SciBERT, and BioBERT models. In details, BertPooler is a pooling layer employed in the BERT (Bidirectional Encoder Representations from Transformers) model. It is utilized in general language comprehension tasks and various Natural Language Processing (NLP) applications. Specifically, it is employed to extract the representation of the token and is commonly used in classification or similar tasks. BertPooler is designed for language modeling and general-purpose tasks. On the other hand, EsmPooler is a specialized pooling layer in the Evoformer (ESM) model. The ESM model is designed to address the specific requirements of biological data, particularly optimized for data types such as protein sequences. EsmPooler is designed to extract representations of proteins and is used in tasks related to protein structure. This layer is specifically developed for biological data processing tasks. In summary, BertPooler is designed as a pooling layer for general language modeling and NLP tasks, whereas EsmPooler is specifically tailored as a pooling layer for biological data, especially protein sequences. Each is effective within its unique context and usage scenarios.

Another significant stage is the usage of attention map in the architecture of transformer models. Attention Map is used specifically to visualize and comprehend the attention mechanism of a transformer model. Models based on the transformer architecture have the capability to focus on specific parts of input data. This focusing is achieved through an attention weighting process, which helps us understand how much attention the model pays to specific features or portions of the input. This weighting process is visualized as a map or matrix, often containing a input sequence and the corresponding attention weights. This map illustrates how much attention the model allocates to each input element when generating a particular output. Higher attention weights indicate that the model gives significant importance to that input, whereas lower weights suggest lower significance. This process is employed in interpreting the decisions of the model, analyzing its internal mechanisms, and understanding why it produces specific outputs.

In conclusion, this study furnishes an effective framework for the classification of anticancer peptides. By delving into the intricate interplay of various parameters, the research not only enhances our understanding of the underlying mechanisms but also provides a substantial contribution to augmenting model performance in this critical domain.

## 5. Experiment Results

Various evaluation metrics are employed to assess the performance of the proposed transformer models. These metrics include accuracy, sensitivity, specificity, Matthew Correlation Coefficient (MCC), Area Under the Curve (AUC), precision, F1 score, and recall, with the results presented as percentages.

- Accuracy: Accuracy represents the ratio of correct predictions to the total data points. In other words, it's the ratio of correctly predicted data points to all data points. It is often used to assess the overall performance of a classification model. If accuracy is high, it indicates the model's ability to make correct predictions.

- Sensitivity: Sensitivity represents the rate at which positive cases are correctly recognized. It calculates the ratio of true positives to the total positive predictions. This metric shows how well the true positive cases are detected. High sensitivity helps reduce false negatives (missing true positives).

- Specificity: Specificity represents the rate at which negative cases are correctly recognized. It calculates the ratio of true negatives to the total negative predictions. This metric indicates how well true negative cases are recognized. High specificity helps reduce false positives (incorrectly labeling true negatives as positive).

- Matthew Correlation Coefficient (MCC): MCC is a value that measures the balance between sensitivity and specificity. It takes values between -1 and +1, with 1 representing the best result. MCC is a balanced measure of a classification model's performance. +1 represents perfect agreement, 0 represents random predictions, and -1 represents complete inverse predictions.

- Area Under the Curve (AUC): AUC represents the area under the ROC (Receiver Operating Characteristic) curve. The ROC curve is a graphical representation of the relationship between sensitivity and specificity. AUC is a metric that evaluates a classification model's class separation ability. As it approaches 1, it is considered that the model has better performance.

- Precision: Precision represents the ratio of data points predicted as positive that are actually positive. In other words, it's the ratio of true positives to the total positive predictions. Precision focuses on reducing false positives and is particularly important in applications where the cost of false positives is high.

- F1 Score: F1 Score represents the harmonic mean of precision and sensitivity. It measures the balanced performance of a classification model. A high F1 Score indicates a model where both precision and sensitivity are high.

- Recall: Recall represents the rate at which true positive cases are correctly recognized. It calculates the ratio of true positives to the total positive cases. Recall focuses on reducing false negatives and is particularly important in applications where positive cases are critical.

ACP-ESM: A novel framework for classification of anticancer peptides using protein-oriented transformer approach

The prediction analysis of the proposed transformers on the Anticp2 Alternative dataset is provided in Table 4. For training, the Anticp2 Alternative internal data is used, and for validation, the Anticp2 alternative validation data is utilized. The model with the highest performance is marked in bold. Performance results are evaluated using different hyperparameters to measure the discriminative power of the classifiers. The highest success is achieved by the ESM model. ESM model achieves an accuracy of 96.45% with the following hyperparameters: 24 epochs, a batch size of 24, a learning rate of 7e-5, gradient accumulation of 2, a warm-up ratio of 0.5, weight decay of 0.0001, and a logging level of 2. In contrast, the ProtBert model yields lower accuracy result compared to the ESM model. ProtBert model achieves an accuracy of 93.00% with 24 epochs, a batch size of 48, a learning rate of 3e-5, gradient accumulation of 2, a warm-up ratio of 0.1, weight decay of 0.001, and a logging level of 4. On the other hand, the performance of the SciBert and BioBert models are less satisfactory compared to the ESM and ProtBert methods.

Table 4: Experiment Results of Transformer Models for AntiCp2 Alternate Dataset

| Models | Epoch | Batch | Learning Rate | Gradient Accumulation Steps | Warmup Ratio | Weight Decay | Logging Steps | Accuracy |
|---|---|---|---|---|---|---|---|---|
| ESM | 24 | 64 | 1e-6 | 2 | 0.3 | 0.0001 | 10 | 0.9278 |
| ESM | 16 | 36 | 1e-5 | 2 | 0.5 | 0.001 | 4 | 0.9304 |
| ESM | 36 | 24 | 5e-6 | 2 | 0.6 | 0.0001 | 2 | 0.9300 |
| ESM | 24 | 90 | 8e-6 | 2 | 0.5 | 0.001 | 2 | 0.9481 |
| **ESM** | **24** | **24** | **7e-5** | **2** | **0.5** | **0.0001** | **2** | **0.9645** |
| ProtBert | 32 | 24 | 6e-6 | 4 | 0.3 | 0.001 | 2 | 0.9119 |
| ProtBert | 24 | 36 | 1e-6 | 2 | 0.1 | 0.001 | 4 | 0.9067 |
| ProtBert | 24 | 48 | 3e-5 | 2 | 0.1 | 0.001 | 4 | 0.9300 |
| ProtBert | 24 | 64 | 5e-5 | 2 | 0.2 | 0.001 | 2 | 0.9196 |
| ProtBert | 24 | 60 | 9e-6 | 2 | 0.3 | 0.001 | 2 | 0.9278 |
| SciBERT | 16 | 24 | 1e-5 | 4 | 0.3 | 0.001 | 10 | 0.8964 |
| SciBERT | 24 | 96 | 1e-6 | 2 | 0.1 | 0.001 | 10 | 0.7387 |
| SciBERT | 36 | 96 | 5e-6 | 2 | 0.1 | 0.001 | 1 | 0.8419 |
| SciBERT | 36 | 156 | 5e-6 | 2 | 0.1 | 0.001 | 4 | 0.8580 |
| SciBERT | 8 | 16 | 1e-4 | 2 | 0.8 | 0.0001 | 10 | 0.8536 |
| BioBERT | 32 | 36 | 1e-5 | 2 | 0.1 | 0.0001 | 4 | 0.9012 |
| BioBERT | 32 | 36 | 1e-6 | 2 | 0.1 | 0.0001 | 4 | 0.8575 |
| BioBERT | 32 | 36 | 3e-6 | 2 | 0.3 | 0.0001 | 4 | 0.9041 |
| BioBERT | 32 | 64 | 8e-6 | 2 | 0.3 | 0.0001 | 2 | 0.9067 |
| BioBERT | 32 | 128 | 6e-6 | 2 | 0.3 | 0.0001 | 1 | 0.9015 |

The prediction analysis on the cACP-DeepGram dataset is presented in Table 5. Lee dataset is utilized for training, while independent dataset is employed for validation. Similar to the Anticp2 dataset, the highest performance on the cACP-DeepGram dataset is achieved by the ESM model. In this model, 36 epochs, a batch size of 24, a learning rate of 3e-5, 1 gradient accumulation, a 0.5 warm-up ratio, 0.0001 weight decay, and 2-level logging were utilized to achieve an accuracy of 97.66%. On the other hand, the ProtBert model yields results similar to the ESM model. For the ProtBert model, 32 epochs, a batch size of 24, a learning rate of 9e-6, 4 gradient accumulation, a 0.3 warm-up ratio, 0.001 weight decay, and 2-level logging are employed, resulting in an accuracy of 97.33%. In contrast, the BioBERT and SciBERT models achieve results that are close to other models but slightly lower. For the BioBERT model, 36 epochs, a batch size of 36, a learning rate of 1e-5, 1 gradient accumulation, a 0.1 warm-up ratio, 0.0001 weight decay, and 2-level logging are used, obtaining an accuracy of 96.67%. Regarding the SciBERT model, 32 epochs, a batch size of 36, a 96% learning rate, 5e-5, 2 gradient accumulation, a 0.05 warm-up ratio, 0.001 weight decay, and 2-level logging are applied, resulting in an accuracy of 96.67%.

Table 5: Experiment Results of Transformer Models for cACP-DeepGram Dataset

| Models | Epoch | Batch | Learning Rate | Gradient Accumulation Steps | Warmup Ratio | Weight Decay | Logging Steps | Accuracy |
|---|---|---|---|---|---|---|---|---|
| ESM | 16 | 24 | 1e-7 | 10 | 0.05 | 0.0001 | 5 | 0.9106 |
| ESM | 40 | 64 | 1e-5 | 8 | 0.5 | 0.0001 | 10 | 0.9700 |
| ESM | 48 | 24 | 5e-6 | 2 | 0.1 | 0.001 | 1 | 0.9567 |
| ESM | 64 | 24 | 8e-6 | 4 | 0.3 | 0.001 | 2 | 0.9600 |
| **ESM** | **36** | **24** | **3e-5** | **1** | **0.5** | **0.0001** | **2** | **0.9766** |
| ProtBert | 16 | 24 | 6e-5 | 2 | 0.6 | 0.001 | 2 | 0.9253 |
| ProtBert | 32 | 24 | 9e-6 | 4 | 0.3 | 0.001 | 2 | 0.9733 |
| ProtBert | 24 | 32 | 1e-4 | 8 | 0.5 | 0.0001 | 10 | 0.7109 |
| ProtBert | 32 | 24 | 2e-5 | 2 | 0.1 | 0.001 | 2 | 0.9500 |
| ProtBert | 16 | 16 | 5e-5 | 2 | 0.5 | 0.001 | 2 | 0.9476 |
| SciBERT | 32 | 24 | 6e-6 | 4 | 0.5 | 0.001 | 2 | 0.9267 |
| SciBERT | 16 | 16 | 1e-4 | 1 | 0.1 | 0.00001 | 5 | 0.6966 |
| SciBERT | 32 | 36 | 1e-5 | 2 | 0.3 | 0.001 | 1 | 0.9567 |
| SciBERT | 32 | 24 | 3e-5 | 2 | 0.5 | 0.0001 | 2 | 0.9633 |
| SciBERT | 32 | 36 | 5e-5 | 2 | 0.05 | 0.001 | 2 | 0.9667 |
| BioBERT | 16 | 24 | 2e-6 | 2 | 0.1 | 0.01 | 2 | 0.9167 |
| BioBERT | 16 | 24 | 8e-6 | 2 | 0.1 | 0.001 | 2 | 0.9455 |
| BioBERT | 24 | 16 | 5e-5 | 4 | 0.4 | 0.0001 | 2 | 0.9325 |
| BioBERT | 36 | 36 | 1e-5 | 1 | 0.1 | 0.0001 | 2 | 0.9667 |
| BioBERT | 64 | 36 | 1e-6 | 1 | 0.05 | 0.00001 | 5 | 0.9533 |

The prediction analysis on the ACP740 dataset is presented in Table 6. The ACP740 dataset is randomly and evenly divided into 80% for training and 20% for validation. Similar to the Anticp2 Alternate and cACP-DeepGram datasets, the ESM model achieves the highest performance on the ACP740 dataset. In this model, 24 epochs, a batch size of 36, a learning rate of 5e-5, 2 gradient accumulations, a warm-up rate of 0.7, a weight decay of 0.001, and 1-level logging are used to achieve an accuracy of 88.51%. On the other hand, Scibert performs 85.81% of accuracy, Biobert provides 84.65% of accuracy, and Protbert achieves 84.45% of accuracy, all of which are slightly lower but close to the performance of the ESM model. In the case of the Scibert model, 16 epochs, a batch size of 24, a learning rate of 2e-5, 2 gradient accumulations, a warm-up rate of 0.3, a weight decay of 0.0001, and 2-level logging are employed to achieve an accuracy of 85.81%.

In Table 4, Table 5, and Table 6 provide information about all hyper-parameter settings, random 5 results for each dataset with the accuracy values out of 50 trials are listed. The parameter details listed in the tables include epoch size, batch size, learning rate, gradient accumulation step, warm-up ratio, weight decay, and logging step.

In Table 7, the classification performance of ESM model is demonstrated using various evaluation metrics. It is obviously observed that the ESM model performs well on different datasets providing consistent results in terms of different evaluation metrics. First, on the cACP-DeepGram dataset, the ESM model achieves impressive results with 97.66% of accuracy, 98.00% of precision, 98.66% of specificity, and MCC of 0.96. Additionally, it exhibits remarkable classification results for other metrics such as 99.31% of precision, 97.62% of F1 score, 96.00% of recall, and a 0.99 AUC. Moreover, the ESM model for cACP-DeepGram dataset has been able to detect positive classes with a sensitivity value of 98.00% and negative classes with a specificity score of 98.66%, respectively. Next, for the Anticp2 Alternate dataset, the ESM model achieves successful results with 96.45% of accuracy, 95.80% precision, 97.00% specificity, and an MCC of 0.92. Precision is 96.478%, the F1 score is 96.14%, recall is 95.80%, and the AUC score is 0.98, demonstrating high values in other performance metrics as well. Furthermore, the ESM model for Anticp2 Alternate dataset has been able to detect positive classes with a sensitivity value of 95.80% and negative classes with a specificity score of 97.00%, respectively. Finally, for the ACP-740 dataset, the model displays satisfactory performance with 88.51% of accuracy, 88.00% of precision, 84.93% of specificity, and an MCC of 0.72. Precision is 88.15%, the F1 score is 88.74%, recall is 89.33%, and the AUC is 0.94, achieving successful results in other metrics as well. In addition, the ESM model for ACP-740 dataset has been able to detect positive classes with a sensitivity value of 88.00% and negative classes with a specificity score of 84.93%, respectively.

Table 6: Experiment Results of Transformer Models for ACP-740 Dataset

| Models | Epoch | Batch | Learning Rate | Gradient Accumulation Steps | Warmup Ratio | Weight Decay | Logging Steps | Accuracy |
|---|---|---|---|---|---|---|---|---|
| ESM | 36 | 24 | 5e-6 | 2 | 0.6 | 0.001 | 1 | 0.8513 |
| ESM | 16 | 30 | 3e-5 | 4 | 0.6 | 0.001 | 10 | 0.8581 |
| ESM | 36 | 24 | 9e-6 | 4 | 0.7 | 0.001 | 2 | 0.8445 |
| ESM | 36 | 16 | 1e-6 | 2 | 0.5 | 0.0001 | 1 | 0.8239 |
| **ESM** | **24** | **36** | **5e-5** | **2** | **0.7** | **0.001** | **1** | **0.8851** |
| ProtBert | 32 | 24 | 6e-6 | 4 | 0.3 | 0.001 | 1 | 0.8108 |
| ProtBert | 24 | 36 | 5e-5 | 2 | 0.2 | 0.0001 | 1 | 0.8175 |
| ProtBert | 24 | 24 | 1e-5 | 4 | 0.5 | 0.001 | 1 | 0.8310 |
| ProtBert | 24 | 24 | 5e-5 | 4 | 0.5 | 0.001 | 5 | 0.8445 |
| ProtBert | 24 | 24 | 1e-6 | 2 | 0.1 | 0.0001 | 1 | 0.8265 |
| SciBERT | 16 | 24 | 9e-6 | 4 | 0.3 | 0.001 | 1 | 0.8445 |
| SciBERT | 16 | 24 | 2e-5 | 2 | 0.3 | 0.001 | 2 | 0.8581 |
| SciBERT | 16 | 16 | 1e-6 | 4 | 0.5 | 0.0001 | 1 | 0.8184 |
| SciBERT | 24 | 36 | 5e-5 | 2 | 0.1 | 0.001 | 1 | 0.8432 |
| SciBERT | 24 | 24 | 8e-5 | 2 | 0.3 | 0.001 | 1 | 0.7934 |
| BioBERT | 32 | 36 | 1e-5 | 2 | 0.1 | 0.001 | 1 | 0.7770 |
| BioBERT | 32 | 36 | 6e-6 | 2 | 0.4 | 0.0001 | 1 | 0.8378 |
| BioBERT | 24 | 24 | 1e-6 | 4 | 0.5 | 0.001 | 1 | 0.8134 |
| BioBERT | 24 | 24 | 5e-6 | 2 | 0.1 | 0.001 | 1 | 0.8245 |
| BioBERT | 24 | 24 | 5e-5 | 2 | 0.1 | 0.0001 | 1 | 0.8465 |

Table 7: The Classification Performance of ESM Model in terms of Various Evaluation Metrics

| Dataset | Methods | Accuracy | Sensitivity | Specificity | MCC | Precision | F1 | Recall | AUC |
|---|---|---|---|---|---|---|---|---|---|
| Anticp2 Alternate | ESM | 96.451 | 95.804 | 97.006 | 0.9285 | 96.478 | 96.140 | 95.804 | 0.98 |
| cACP-DeepGram | ESM | 97.666 | 98.000 | 98.666 | 0.9605 | 99.310 | 97.627 | 96.000 | 0.99 |
| ACP-740 | ESM | 88.513 | 88.000 | 84.931 | 0.7298 | 88.157 | 88.741 | 89.333 | 0.94 |

Table 8, Table 9, and Table 10 present a comprehensive comparison of the proposed transformer models with the existing literature studies and provide the experimental results. These tables list various performance metrics of the models, including accuracy, sensitivity, specificity, and MCC. In Table 8, noteworthy results for the ESM model on the AntiCp2 Alternate dataset are highlighted. The model demonstrates a remarkable performance with an accuracy rate of 96.45% outperforming the state-of-the-art (SOTA) results. Additionally, it has achieved high results such as a sensitivity score of 95.80%, specificity of 97.00%, and an MCC value of 0.9285. These results emphasize the effectiveness and reliability of the ESM model in handling the AntiCp2 Alternate dataset. In Table 9, a comparison has been made with models found in the literature that use the cACP-DeepGram dataset as their data source. The ESM model achieved very high results, with an accuracy of 97.66%, sensitivity of 98.00%, specificity of 98.66%, and an MCC of 0.9605. In Table 10, a comparison has been made between the ESM model and the ACP-MHCNN model, which is found in the literature and uses the Acp740 dataset. In this comparison, the ESM model has demonstrated superior performance, achieving an accuracy of 88.51%, sensitivity of 88.00%, specificity of 84.93%, and an MCC of 0.7298.

Table 8: Comparison with SOTA for AntiCp2 Alternate Dataset

ACP-ESM: A novel framework for classification of anticancer peptides using protein-oriented transformer approach| Study | Models | Accuracy | Sensitivity | Specificity | MCC |
|---|---|---|---|---|---|
| Chen et al. (2016) | iACP | 77.6 | 78.4 | 76.8 | 0.55 |
| Wei et al. (2019) | PEPred-Suite | 57.5 | 40.2 | 74.7 | 0.16 |
| Rao et al. (2020) | ACPpred-Fuse | 78.9 | 64.4 | 93.3 | 0.60 |
| Wei et al. (2018b) | ACPred-FL | 43.8 | 60.2 | 25.6 | -0.15 |
| Schaduangrat et al. (2019) | ACPred | 85.3 | 87.1 | 83.5 | 0.71 |
| Tyagi et al. (2013) | AntiCP | 90.0 | 89.7 | 90.2 | 0.80 |
| Agrawal et al. (2021) | AntiCP_2.0 | 92.0 | 92.3 | 91.8 | 0.84 |
| Lv et al. (2021) | iACP-DRLF | 93.0 | 89.6 | 96.4 | 0.86 |
| **Proposed Method** | **ESM** | **96.45** | **95.80** | **97.00** | **0.9285** |

Table 9: Comparison with SOTA for cACP-DeepGram Dataset

| Study | Models | Accuracy | Sensitivity | Specificity | MCC |
|---|---|---|---|---|---|
| Manavalan et al. (2017) | SVMACP | 81.40 | 77.50 | 85.30 | 0.63 |
| Manavalan et al. (2017) | RFACP | 82.70 | 70.60 | 94.80 | 0.67 |
| Ahmed et al. (2021) | ACP-MHCNN | 78.9 | 88.90 | 83.10 | 0.72 |
| Feng et al. (2022) | ME-ACP | 89.70 | 87.50 | 91.80 | 0.80 |
| Boopathi et al. (2019) | mACPpred | 91.40 | 88.50 | 94.30 | 0.83 |
| He et al. (2021) | ACPred-LAF | 93.27 | 89.07 | 92.88 | 0.87 |
| Akbar et al. (2022) | cACP-DeepGram | 96.94 | 97.67 | 89.10 | 0.90 |
| **Proposed Method** | **ESM** | **97.66** | **98.00** | **98.66** | **0.9605** |

Table 10: Comparison with SOTA for ACP-740 Dataset

| Study | Models | Accuracy | Sensitivity | Specificity | MCC |
|---|---|---|---|---|---|
| Ahmed et al. (2021) | XGB | 81.6 | 82.4 | 81.8 | 0.64 |
| Ahmed et al. (2021) | ET | 81.5 | 78.4 | 85.9 | 0.65 |
| Ahmed et al. (2021) | RF | 81.2 | 79.2 | 84.8 | 0.64 |
| Ahmed et al. (2021) | ACP-DL | 80.0 | 81.4 | 78.6 | 0.60 |
| Ahmed et al. (2021) | ACP-MHCNN | 86.0 | **88.9** | 83.1 | 0.72 |
| **Proposed Method** | **ESM** | **88.51** | 88.00 | **84.93** | **0.7298** |

To demonstrate the computational efficiency of the proposed model, the performance of each model on two different dataset is presented in terms of training and inference time in Table 11. The obtained results indicate a notable advantage of the ESM model in both training and inference durations compared to the other models. To comprehend the superior performance of the ESM model, it is hypothesized that the model's unique design and learning algorithms are able to align well with the specific characteristics of the dataset. The enhanced capability of ESM to represent protein sequences more effectively allows it to learn the properties of anticancer peptides more accurately. Additionally, the rapid learning ability of ESM contributes to optimizing training times and achieving impressive speed during inference. These findings shed light on the reasons behind the preference for the ESM model in anticancer peptide classification applications. The model's quick learning proficiency and effective representation power make it particularly well-suited for this specific task.

As obviously observed from the Fig. 3, the ESM model, which we advocate for use across all datasets, possesses a modest parameter count, and this simplicity contributes positively to its performance. The model's limited number of parameters enhances its ability to generalize effectively, mitigating the risk of over-fitting. Consequently, this characteristic aids the model in achieving favorable outcomes across a diverse array of data. Additionally, the model's simplicity necessitates less memory and processor resources, rendering it a more efficient choice for application and

Kilimci and Yalcin: *Preprint submitted to Elsevier* Page 17 of 20

deployment. Hence, the proposed model strikes a commendable balance, characterized by its restrained parameter count and superior performance.

Table 11: Computational Performances of the Transformer Models on Each Dataset

| Models | Dataset | Training Time | Inference Time (in seconds) |
| --- | --- | --- | --- |
| ESM | AntiCp2 Alternate | 3m 47s | 0.2370 |
| ESM | cACP-DeepGram | 2m 41s | 0.2145 |
| ESM | Acp740 | 1m 8s | 0.2568 |
| ProtBert | AntiCp2 Alternate | 10m 42s | 0.2627 |
| ProtBert | cACP-DeepGram | 10m 10s | 0.2899 |
| ProtBert | Acp740 | 4m 21s | 0.2780 |
| SciBERT | AntiCp2 Alternate | 4m 31s | 0.2536 |
| SciBERT | cACP-DeepGram | 6m 29s | 0.2983 |
| SciBERT | Acp740 | 3m 4s | 0.2875 |
| BioBERT | AntiCp2 Alternate | 8m 26s | 0.2705 |
| BioBERT | cACP-DeepGram | 7m 45s | 0.2734 |
| BioBERT | Acp740 | 5m 4s | 0.2682 |

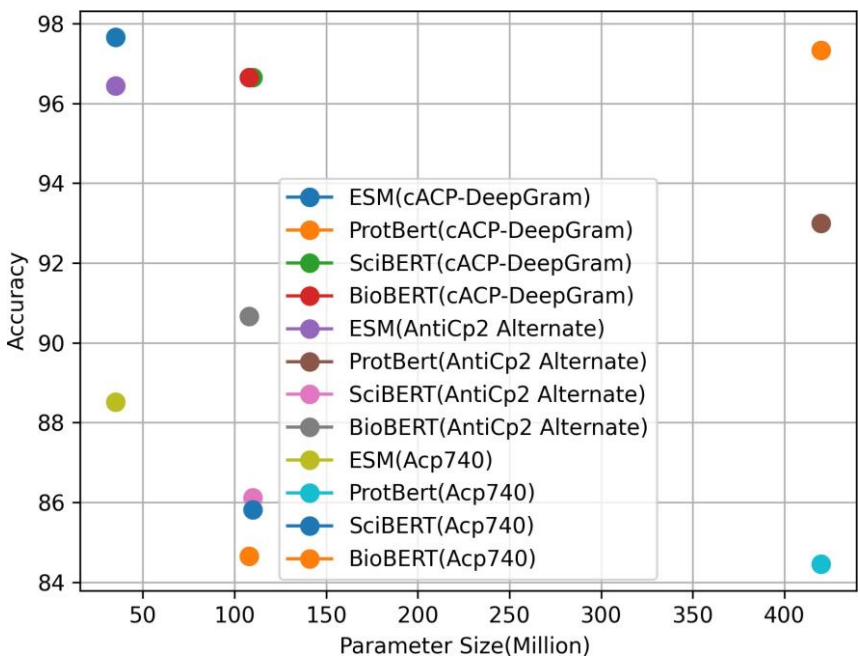

**Figure 3:** Evaluation of Parameter Size and Classification Performance for All Models on Each Dataset

## 6. Discussion and Conclusion

Our study explores the potential of anticancer peptides (ACPs) as a promising avenue in cancer research and therapy. The unique ability of ACPs to selectively target and eliminate cancer cells, while preserving healthy cells to a considerable extent, positions them as valuable candidates for innovative cancer treatments. Despite their promise,

ACP-ESM: A novel framework for classification of anticancer peptides using protein-oriented transformer approachthe practical application of ACPs faces challenges such as stability optimization, improved selectivity, and enhanced delivery to cancer cells. Additionally, the increasing number of peptide sequences demands the development of reliable and precise prediction models. In response to these challenges, our work presents an efficient transformer-based framework designed to identify anticancer peptides accurately. Leveraging four distinct transformer models, namely, ESM, ProtBert, BioBERT, and SciBERT, we conduct experiments on well-established datasets, including AntiCp2, cACP-DeepGram, and ACP-740. The results underscore the efficacy of our proposed framework, particularly the ESM model, which achieved remarkable accuracy rates of 96.45%, 97.66%, and 88.51% for the AntiCp2, cACP-DeepGram, and ACP-740 datasets, respectively. These outcomes position our framework as a new benchmark in the field, outperforming existing studies.

The success of our proposed framework, notably the ESM model's superior performance, underscores the potential of transformer-based approaches in accurately identifying anticancer peptides. The high classification accuracy achieved across diverse datasets supports the robustness and generalizability of our model. The ability to surpass state-of-the-art studies in terms of accuracy marks a significant advancement, emphasizing the practical applicability of our framework in real-world scenarios. However, challenges persist, and future work should focus on addressing these issues. Optimization of ACP stability, enhancement of selectivity, and improvement in delivery mechanisms remain critical areas for further investigation. Additionally, the continuous expansion of peptide sequences necessitates ongoing efforts to refine and expand prediction models. In summary, our study contributes a robust transformer-based framework that not only outperforms existing models but also sheds light on the immense potential of ACPs in cancer therapeutics. As we move forward, collaborative efforts within the scientific community will be crucial to overcoming the remaining challenges and advancing ACPs from a promising concept to a practical and effective cancer treatment modality.

## 7. Declaration of Generative AI and AI-Assisted Technologies in the Writing Process

During the preparation of this work the authors used ChatGPT tool in order to improve language and readability. After using this tool, the authors reviewed and edited the content as needed and takes full responsibility for the content of the publication.

## References

Agrawal, P., Bhagat, D., Mahalwal, M., Sharma, N., Raghava, G.P., 2021. Anticp 2.0: an updated model for predicting anticancer peptides. Briefings in bioinformatics 22, bbaa153.

Ahmed, S., Muhammod, R., Khan, Z.H., Adilina, S., Sharma, A., Shatabda, S., Dehzangi, A., 2021. Acp-mhcnn: An accurate multi-headed deep-convolutional neural network to predict anticancer peptides. Scientific reports 11, 23676.

Akbar, S., Hayat, M., Tahir, M., Khan, S., Alarfaj, F.K., 2022. cacp-deepgram: classification of anticancer peptides via deep neural network and skip-gram-based word embedding model. Artificial intelligence in medicine 131, 102349.

Alimirzaei, F., Kieslich, C.A., 2023. Machine learning models for predicting membranolytic anticancer peptides, in: Computer Aided Chemical Engineering. Elsevier. volume 52, pp. 2691–2696.

Alsanea, M., Dukyil, A.S., Afnan, Riaz, B., Alebeisat, F., Islam, M., Habib, S., 2022. To assist oncologists: An efficient machine learning-based approach for anti-cancer peptides classification. Sensors 22, 4005.

Ammar, W., Groeneveld, D., Bhagavatula, C., Beltagy, I., Crawford, M., Downey, D., Dunkelberger, J., Elgohary, A., Feldman, S., Ha, V., et al., 2018. Construction of the literature graph in semantic scholar. arXiv preprint arXiv:1805.02262 .

Beltagy, I., Lo, K., Cohan, A., 2019. Scibert: A pretrained language model for scientific text. arXiv preprint arXiv:1903.10676 .

Boopathi, V., Subramaniyam, S., Malik, A., Lee, G., Manavalan, B., Yang, D.C., 2019. macppred: a support vector machine-based meta-predictor for identification of anticancer peptides. International journal of molecular sciences 20, 1964.

Chen, J., Cheong, H.H., Siu, S.W., 2021. xdeep-acpep: deep learning method for anticancer peptide activity prediction based on convolutional neural network and multitask learning. Journal of chemical information and modeling 61, 3789–3803.

Chen, W., Ding, H., Feng, P., Lin, H., Chou, K.C., 2016. iacp: a sequence-based tool for identifying anticancer peptides. Oncotarget 7, 16895.

Deng, H., Ding, M., Wang, Y., Li, W., Liu, G., Tang, Y., 2023. Acp-mlc: A two-level prediction engine for identification of anticancer peptides and multi-label classification of their functional types. Computers in Biology and Medicine 158, 106844.

Elnaggar, A., Heinzinger, M., Dallago, C., Rehawi, G., Wang, Y., Jones, L., Gibbs, T., Feher, T., Angerer, C., Steinegger, M., et al., 2021. Prottrans: Toward understanding the language of life through self-supervised learning. IEEE transactions on pattern analysis and machine intelligence 44, 7112–7127.

Fazal, E., Ibrahim, M.S., Park, S., Naseem, I., Wahab, A., 2023. Anticancer peptides classification using kernel sparse representation classifier. IEEE Access 11, 17626–17637.

Feng, G., Yao, H., Li, C., Liu, R., Huang, R., Fan, X., Ge, R., Miao, Q., 2022. Me-acp: Multi-view neural networks with ensemble model for identification of anticancer peptides. Computers in Biology and Medicine 145, 105459.Kilimci and Yalcin: *Preprint submitted to Elsevier* Page 19 of 20